\documentclass[12pt]{article}
\begin{document}


\centerline{\bf Application of Tuncay's language teacher model}
\medskip
\centerline{\bf to business-customer relations}

\bigskip
Carmen Costea

\bigskip
Academy of Economic Studies, Bucharest, Romania, 

\bigskip
e-mail:cecostea@ase.ro 
\bigskip

Abstract: {\small
It seems that what has been said by now about market and competitiveness do not 
fit perfectly with competences of getting the best of profit. Sometimes, the 
classical methods of fundamentals of management do not apply to individual 
companies that face irregular accommodation on the market. It is high time to 
replace the perfect business with the right one. New approaches and models may 
help in identifying new competition trends, changes for better application of 
purposes and proposals.}

Key words: complex systems, information, knowledge, business intelligence risk, 
management

\section{A point of view}
The need to build the right business requires the study and understanding of both the business as a complex matter and of the framework established inside the nature and society. The complex approaches of life and profit, of goals and wishes, of objectives and targets are searching for customized ways of operations and behavior to be considered, new ideas to sustain explanations, new tools to be used. It should be that a state-of-the-art knowledge bridge is able to carry across the gap between natural, social and economic environment and build the right world in a society that wants to be proud of it. 

The social and economic importance of life and business complexity and their link with high standards and profit is, sometimes, risky. Under such circumstances, unexpected demands are brought out when growing global stands for higher integration of business processes and improved efficiency, sometimes unfortunately connected to failures or bankruptcy when exposed to unknown risks. 

As people live inside the society their choices are affected by opinions and decisions taken at general level by partners and other agents such as, media and opinion leaders \cite{comment2}. To get the best of their decisions people do network and this way, they contribute to a new life-skill. Everything is moving on inside networks. Ignoring the network - presumably they do not network only when they want something - will not simplify the meaning of the fundamental aspects of life at any level or degree. The networking world is open to anyone, as long as the network values are strong and ethical for their members.  

\section{When less is more and better is bigger: old techniques or new tools}

Looking beyond the words, understanding the body language of individuals or groups, improving the listening skills generate no-boundaries communication, where everybody is somebody, where any question asked will get an answer to clarify details, where the knowledge limits are moved on as there is always more to know. 

To avoid additional risks some orientations are underlined to be considered:
\begin{itemize}
\item{the identification of all possibilities of gains by wide increased efficiency;}
\item{transparent functionality;}
\item{high quality performance;}
\item{integrated approach of both shops/companies and clients inside the same complex system;}
\item{rethinking of consolidation of extension of the partnership with any client.} 
\end{itemize}

In other words, smart companies realize that corporate customers are more knowledgeable than one might think, and consequently seek knowledge through direct interaction with customers, in addition to seeking knowledge about customers from their sales representatives. 

Customer knowledge can be approached under two aspects. The first one runs as a ``collection of information and viewpoints that an organization has about its customers''. The second step takes into consideration the axiom that customers \cite{devries}  are ``more knowledgeable'' than we realize. Thoughts, behaviours, wishes, habits, actions are always important fields providing new information. While knowledge provides the skill to create \cite{devries} new things, intelligence, relying on smart analysts to interpret and communicate unique insights, assign their implications to decision-makers and policy makers \cite{sawka}. 

Customer knowledge should include: gaining, sharing and expanding knowledge of (inside) the customer.  Individual or group experiences in applications, competitor behaviour, possible future solutions together with quantitative insights, a wider variety of less structured information as their own ways of gaining, sharing and expanding knowledge with other customers that will help to build insights into customer relationships. It should include information about: 
\begin{itemize}
\item{individuals (emotional too),}
\item{customer experience and creativity,}
\item{loyalty schemes,}
\item{active lobbying knowledge and knowledge partner,}
\item{customer success,}
\item{innovation,}
\item{organizational learning, performance against budget; customer retention rate; performance in terms of customer satisfaction.}
\end{itemize}
 That helps explaining who those individuals are, what they do and what they are looking for, and may enable broader analysis of customer base as a whole. 

The new aim of companies should be to setting up a strong systemic network of customers so that they can build and manage customer knowledge and relationships over longer terms. To bring some help a new application of the use of Formation of Languages; Equality, Hierarchy and Teachers models is suggested \cite{tuncay}. In this respect, we propose to use the above mentioned model - developed by \c{C}aglar Tuncay \cite{tuncay} - into another application where languages are replaced with $N$ supermarkets brands with a different number of shops spread into a certain geographical area. Each shop has $k$ real customers with $M$ needs ($k\leq N$, $i\leq  M$). For any need there may exist many related items assortments or varieties belonging to different classes of products $j$ underlining, symbolically, five different subentries at most. So we take $1 \leq j \leq j_{\max}$ for every need $w$ and for each $j$ we assign a representative real number $r$. The maximum number $j_{max}$ of subentries $r$ is also determined randomly between 1 and 5, independently for any need $w$. Clearly, $r_{kij} = 0$ ($w_{ki}$ = \{0, 0, 0, 0, 0\}) corresponds to an unknown need of the customer $k$. Customers' needs are groups of up to five real numbers: $w_{ki} = r_{kij}$. (1) \cite{tuncay}

Initially there is no consensus about purchases, but the consensus may be set through several processes. The brand will emerge in that area with the number of customers walking to a certain shop to shop. Their wish $Lk$ varies from customer to customer, mainly at the opening of the shop and this fluctuation fades down with time since $Lk \to L$, if convergence occurs. Later, by assigning to each customer a random real number (rank; greater than or equal to zero, and less than one) a hierarchy may be established. Some leaders opinions with ultimate rank of unity becoming decisions to choose a certain brand should also bring some more information. 

Applying the model, the values for initial $r_{kij}$'s should be changed into time dependent ones, i.e. $r_{kij}(t)$. The purpose is to check the customers' behaviour and the way they look for satisfying their needs and buy from one place or another, from one hypermarket brand or the competitors', on their own need or under the impulse of opinion's leader. 

This model may bring interesting information on what they buy, when they buy, why and how much for. In the long term the company may think to design new products with more added value incorporated in order to satisfy those needs issued from emotional behaviours. This way \cite{tofler}, customers will fill the dual role of both producer and consumer in creating net value inside a team-based co-learning process where mutual innovation and joint intellectual property are obviously considered. There is no point in thinking wealth and durable profit underestimating the customers diversity and falling into the possible ``trap'' of over-reliance on (existing) customer knowledge, without appropriate sensing of wider environmental impacts and influences.  

Mutual understanding, reliance and confidentiality must be agreed upon and consistently implemented. Careful consideration has to be given to degrees of openness in sharing of knowledge, and cultural issues of respect, trust and ways of interacting have to be adequately co-shaped and optimized.

It is quite a challenge to build successful models of a wide range of socio-economic phenomena in which the agents operate with low cognition, both in their ability to gather information and process it. This is in direct contrast to the high levels of cognition assigned in economic theory, even under bounded rationality. A key feature of all these models is that agents operate on networks. A particular interest is the extent to which assigning greater cognition to agents means that models become successful in terms of their dynamic consistency with stylized facts. In order to model natural human behaviour, it is firstly necessary to capture this behaviour.

  The start consists in modelling behaviour for specific situations; once human behaviour captured, the following experiments should be performed. Given a virtual environment, a sufficient number of subjects are asked to execute a human task in this virtual environment. The hypothesis is that the combination of the motion paths and the clues for making/changing decisions will provide decision rules to make reliable predictions about human behaviour under the same conditions when using virtual persons. Later, if the traditional definition of customer is expanding, other parties such as suppliers can also be included. They may be another rich source of valuable knowledge for companies, a source that lies beyond the usual corporate boundaries.

\end{document}